\documentclass[aps,pra,twocolumn,superscriptaddress,showpacs]{revtex4}
\usepackage{epsfig}
\usepackage{amssymb,amsmath}
\begin{document}

\title{Entanglement discontinuity}

\author{Amir Kalev}
\affiliation{Centre for Quantum Technologies, National University of Singapore, 3 Science Drive 2, 117543, Singapore}
\author{Faqir C. Khanna}
\affiliation{Department of physics, University of Alberta, Edmonton,
Alberta, Canada T6G 2J1}
\author{Michael Revzen}
\affiliation{Department of Physics, Technion - Israel Institute of Technology, Haifa
32000, Israel}

\date{\today}

\begin{abstract}
We identify a class of two-mode squeezed states which are parametrized by an angular variable ${0\le\theta<2\pi}$ and a squeezing parameter $r$. We show that, for a large squeezing value, these states are either (almost) maximally entangled or product states depending on the value of $\theta$. This peculiar behavior of entanglement is unique for infinite dimensional Hilbert space and has  consequences for the entangling power of unitary operators in such systems. Finally, we show that, at the limit ${r\rightarrow\infty}$ these states demonstrate a discontinuity attribute of entanglement.
\end{abstract}

\maketitle

\section{Introduction}\label{sec:intro}
Quantum Mechanics hoards intriguing variety of phenomena. Many of which hinges on entanglement. A striking example is the teleportation \cite{bennett93} which has no real counterpart in classical physics. Indeed entanglement is at the heart of quantum mechanics.  Its fundamental role was already recognized in 1935  by Einstein, Podolsky and Rosen (EPR) \cite{epr} and by Schr\"{o}edinger \cite{sch}. Schr\"{o}edinger calls it ``the characteristic trait of quantum mechanics'' \cite{sch}. Entanglement, its characterization and properties have been studied ever since (recent reviews are given, e.g. in \cite{review}).  

The  primitive system where entanglement can be defined, is a bipartite system. In such a system, it is intuitively clear that entanglement is attained by two particles (subsystems) in a pure state with nontrivial Schmidt decomposition \cite{peres,knight}. With this understanding in mind, an immense effort has been undertaken to find a measure for entanglement (entanglement monotone). In fact for bipartite systems a standard quantification exists.
For pure states, the von Neumann entropy of either of the two parties, dubbed as entanglement entropy, was shown to be a successful measure of entanglement \cite{pop}. In fact, the entanglement entropy can be computed from a basic feature of a state -- the two-particle concurrence \cite{hill97,wootters98}. While the concurrence itself is intimately related to the `impurity' of the one-particle reduce state (${1-{\rm tr}[\rho^2]}$). 
Therefore the impurity of the reduce state  has also been used as a measure of bipartite entanglement \cite{purity1,purity2,purity3,purity4,purity5}. 

We note here that there is a subtle difference between finite and infinite dimensional Hibert spaces concerning bipartite pure states. In finite dimensional Hilbert space bipartite pure states range from product states to maximally entangled states. However in continuous variable systems pure states could lay outside the Hilbert space and may be treated as an ideal, limit case  of physical states. For example, the EPR  is a maximally entangled state which lays outside the Hilbert space. It can be considered as the infinite squeezing limit of a two-mode squeezed state (TMSS). The latter is a physical state and for any finite non-zero value of squeezing is a non-maximally entangled state.

Except for this subtle point, however, the full range of bipartite entangled states are successfully quantified.  It is therefore interesting to ask the following question: Is entanglement analytical? In other words, suppose we are given a family of states which is parametrized by some physical parameter, if we continuously change this parameter, does the entanglement of the parametrized state change continuously? Similar questions have been addressed \cite{plenio,enk} in the settings of continuous variable systems. There, a peculiar behavior of entanglement,  has been noticed, and have been shown to have both theoretical and practical consequences. From theoretical point of view, it was shown \cite{plenio} that  in  any  neighborhood of  every  product state  lies  an
arbitrarily strongly entangled state. This in turn, for example,  implies on ways to
retain meaningful measures of entanglement in continuous variable systems. To the best of our knowledge, in the literature there is only one explicit example \cite{enk} for this kind of peculiar behavior of entanglement in infinite dimensions. In \cite{enk}, a standard nonlinear optics interaction followed by a  simple interaction with a beam  splitter  was shown to generate an arbitrarily large amount of entanglement in  an  arbitrarily  short  time. This important result has its effect upon the entangling  power of unitary operations. Here, we provide another example for such power.

The example is given in terms of a family of TMSS. We first identify a class of TMSS which are parametrized by an angular variable ${0\le\theta<2\pi}$ and a squeezing parameter $r$, ${\psi(\theta,r)}$. Actually as we shall see it would be enough to consider the example only for ${0\le\theta<\pi}$. The two parameters set up a sequence of states, spreading from product states (${\theta=\pi/2}$ or ${r=0}$) to maximally entangled states (${\theta\neq\pi/2}$ and ${r\rightarrow\infty}$).  Then we calculate the measure of entanglement of a generic state in the family. We show that as the parametrized state becomes more squeezed (${r>>1}$),  an abrupt change in the entanglement happens for certain values of  $\theta$. The entanglement is nearly a constant (of the order of $1$) for every value of $\theta$, except at the vicinity of ${\theta=\pi/2}$  where the entanglement rapidly decreases to zero. At ${\theta=\pi/2}$ the state is an exact product state for every value of $r$. This change is discontinuous in the limit  ${r\rightarrow\infty}$. The same argument will hold for ${\theta=3\pi/2}$ as well.

The paper is organized as follows: We begin, in Sec.~\ref{sec:2} with a particular parametrization of one-particle mutual unbiased bases (MUB) \cite{wootters1,klimov1}, which we have considered earlier \cite{dv}. The parametrization is given in terms of an angular variable ${0\leq\theta< \pi}$. The bases are constructed by considering infinitely squeezed states, in as much as position and momentum eigenstates are infinitely squeezed  unbiased states. This part sets the stage for constructing, in Sec.~\ref{sec:3}, parametrized bipartite maximally entangled states (again with an angular variable $\theta$) which at the end would demonstrate the discontinuous property of entanglement. To get better understanding on how the discontinuity comes about, we apply our parametrization to two-mode squeezed states. Our study compares the entanglement of two  states with close values of $\theta$. We find that at the vicinity of ${\theta=\pi/2}$ there is a rapid change in the entanglement. This change becomes discontinuous in the limit of maximally entangled states. This is done in Sec.~\ref{sec:4}. Finally, we conclude with final remarks and close with a short discussion on a possible experiment to demonstrate this feature of entanglement.

\section{Quadrature Mutually Unbiased Bases}\label{sec:2}

Consider the complete orthonormal basis ${\{|y,\theta \rangle\}_{y\in\mathbb R}}$, ${0\le\theta < 2\pi}$, defined by the eigenvalue equation,
\begin{equation}\label{cos-sin}
(\cos\theta\, \hat{x}+\sin\theta \,\hat{p})\,|y,\theta \rangle \equiv
\Lambda(\theta)|y,\theta \rangle=y\,|y,\theta \rangle.
\end{equation}
This defines ${\Lambda(\theta)}$. ${\Lambda(\theta)}$ has a clear physical meaning, in quantum optics for example, the `position'  and  `momentum'  operators (quadratures) $\hat{x}$  and $\hat{p}$  represent the in-phase and the out-of-phase components of an electric field amplitude with respect to a  strong (classical) reference field ${\propto\,\cos(\theta)}$.  
Note that ${\Lambda(0)=\hat{x},\,\Lambda(\frac\pi{2})=\hat{p}}$, so that ${|y,0 \rangle\equiv|x \rangle}$ is a position eigenstate while ${|y,\frac\pi{2} \rangle\equiv|p \rangle}$ is a momentum eigenstate. It is convenient to consider the operators \cite{ulf},
\begin{equation}\label{aop}
\hat{a}=\frac{1}{\sqrt 2}(\hat {x}+i\hat{p});\,\hat{a}^{\dagger}=\frac{1}{\sqrt 2}
(\hat {x}-i\hat{p});\,{\rm and}\,\, U(\theta)=e^{-i\theta \hat{a}^{\dagger}\hat{a}},
\end{equation}
such that,
\begin{equation}\label{lam}
\Lambda(\theta)=U^{\dagger}(\theta)\hat{x}U(\theta).
\end{equation}

Now, the state ${|y,\theta\rangle}$  may be expressed in terms of the state at ${\theta = 0}$, 
\begin{equation}\label{prels}
|y,\theta \rangle=U^{\dagger}(\theta)|y,0\rangle.
\end{equation}
This defines our phase choice \cite{oded}. (It differs from the standard one \cite{wootters1}.) Thus we may read off the position-representative
solutions for the harmonic oscillator (${m=\omega=1}$) \cite{larry},
\begin{equation} \label{ls}
\langle x|y,\theta \rangle=\frac{1}{\sqrt{2\pi \sin \theta}}
e^{-\frac{i}{2\sin\theta}\left([y^2+x^2]\cos \theta-2yx\right)}.
\end{equation}
Two bases (in our case, two distinct values for $\theta$) are said to be MUB if and only if the {\it
magnitude} of the scalar product of a vector belonging to one basis with one
belonging to the other basis is independent of their vectorial (intra basis) labels.
We  verify that the bases labelled  $\theta$ and $\theta'$ where ${\theta \ne \theta'}$,
are MUB:
\begin{equation}\label{mub}
|\langle y', \theta '|y, \theta \rangle|=|\langle
x'|U^{\dagger}(\theta-\theta')|x\rangle|=\frac{1}{\sqrt{2\pi \sin(\theta-\theta')}}.
\end{equation}
We now define the state ${|k,\theta \rangle}$ as
\begin{equation}
(-\sin\theta\,\hat{x}+\cos\theta\,\hat{p})\,|k,\theta \rangle=k\,|k,\theta \rangle.
\end{equation}
Then, similarly to Eqs.~(\ref{lam},\ref{prels})
\begin{equation}
U^{\dagger}(\theta)\hat{p}U(\theta)|k,\theta \rangle=k\,|k,\theta \rangle,
\end{equation}
and
\begin{equation}\label{p}
|k,\theta \rangle=U^{\dagger}(\theta)|k,0\rangle.
\end{equation}
Here, ${|k,0 \rangle\equiv|p \rangle}$ is a momentum eigenstate.
Now the well known Fourier transform relation implies,
\begin{equation}\label{xp}
\langle y,\theta|k,\theta \rangle=\frac{1}{\sqrt{2\pi}}e^{iky}.
\end{equation}
We note that, since ${|y,\frac{\pi}{2}\rangle}$ is an
eigenfunction of $\hat{p}$, with eigenvalue $y$, we have,
\begin{equation}\label{pxrel}
\langle y,\frac{\pi}{2}|k,0\rangle=\delta(y-k),
\end{equation}
underscoring that ${|y,\frac{\pi}{2}\rangle}$ is a momentum eigenstate with an eigenvalue $y$, ${\hat{p}\,|y,\frac{\pi}{2}\rangle=y\,|y,\frac{\pi}{2}\rangle}$, and ${|k,0\rangle}$ is also a momentum eigenstate with an eigenvalue $k$, \mbox{$\hat{p}\,|k,0\rangle=k\,|k,0\rangle$}. The $\theta$ labelling of MUB suggests its extension to two-particle entangled states.

\section{Entangled Quadratures}\label{sec:3}

The generic entangled state in phase space (i.e. no spin) is the EPR state
\begin{equation}\label{epr}
|\xi,\mu \rangle=\frac{1}{\sqrt{2\pi}}\int dx_1dx_2\delta \left( \frac{x_1-x_2}{\sqrt 2}-\xi \right)e^{i \mu \frac{x_1+x_2}{\sqrt2}} |x_1\rangle |x_2\rangle.
\end{equation}
This state is an eigenstate of the commuting operators,
\begin{equation}\label{ximu}
\hat
{\xi}=\frac{1}{\sqrt2}(\hat{x}_1-\hat{x}_2),\,\,\hat{\mu}=\frac{1}{\sqrt2}(\hat{p}_1+\hat{p}_2).
\end{equation}
These operators when combined with
\begin{equation}
 \hat
{\eta}=\frac{1}{\sqrt2}(\hat{x}_1+\hat{x}_2),\,\,\hat{\nu}=\frac{1}{\sqrt2}(\hat{p}_1-\hat{p}_2)
\end{equation}
\noindent form a complete set of operators.
By analogy with our analysis above we now
consider the complete orthonormal bases $\{|\xi,\theta \rangle|\eta,\theta'
\rangle\}$ defined as,
\begin{eqnarray}
(\cos\theta \,\hat{\xi}+\sin\theta\, \hat{\nu})|\xi,\theta \rangle&=&
\xi\,  |\xi,\theta \rangle, \nonumber\\
(\cos\theta \,\hat{\eta}+\sin\theta \,\hat{\mu})|\eta,\theta \rangle&=&
\eta\, |\eta,\theta \rangle.
\end{eqnarray}
Alternatively,  we may consider the orthonormal bases ${\{|\mu,\theta \rangle|\nu,\theta'
\rangle\}}$ defined as,
\begin{eqnarray}\label{EntBases}
(-\sin\theta \,\hat{\xi}+\cos\theta\, \hat{\nu})|\nu,\theta \rangle&=&
\nu\,  |\nu,\theta \rangle, \nonumber\\
(-\sin\theta \,\hat{\eta}+\cos\theta \,\hat{\mu})|\mu,\theta \rangle&=&
\mu\, |\mu,\theta \rangle.
\end{eqnarray}
We define the following operators
\begin{eqnarray}
\hat{A}&=&\frac{1}{\sqrt2}(\hat{\xi}+i\hat{\nu}),\,\,\hat{A}^{\dagger}=\frac{1}{\sqrt2}(\hat{\xi}-i\hat{\nu}),\nonumber\\
\hat{B}&=&\frac{1}{\sqrt2}(\hat{\eta}+i\hat{\mu}),\,\,\hat{B}^{\dagger}=\frac{1}{\sqrt2}(\hat{\eta}-i\hat{\mu}),
\end{eqnarray}
that obey the commutation relations
\begin{equation}
\left[\hat{A},\hat{A}^{\dagger}\right]=\left[\hat{B},\hat{B}^{\dagger}\right]=1,
\end{equation}
with all other commutators vanishing. We define
\begin{equation}
V_A(\theta)=e^{-i\theta \hat{A}^{\dagger}\hat{A}};\,\,V_B(\theta)=e^{-i\theta
\hat{B}^{\dagger}\hat{B}}.
\end{equation}
With these definitions, we may write
\begin{eqnarray}
V_A^{\dagger}(\theta)\,\hat{\xi}\,V_A(\theta)|\xi,\theta\rangle&=&\xi\,|\xi,\theta
\rangle,\nonumber\\V_A^{\dagger}(\theta)\,\hat{\nu}\,V_A(\theta)|\nu,\theta\rangle&=&\nu\,|\nu,\theta\rangle,
\end{eqnarray}
and,
\begin{eqnarray}
V_B^{\dagger}(\theta)\,\hat{\eta}\,V_B(\theta)|\eta,\theta \rangle&=&\eta\,|\eta,\theta
\rangle,\nonumber\\
V_B^{\dagger}(\theta)\,\hat{\mu}\,V_B(\theta)|\mu,\theta
\rangle&=&\mu\,|\mu,\theta \rangle.
\end{eqnarray}
It follows that
\begin{eqnarray}\label{vb}
|\xi,\theta \rangle&=&V_A^{\dagger}(\theta)|\xi \rangle,\,\,|\nu,\theta \rangle=V_A^{\dagger}(\theta)|\nu \rangle, \nonumber \\
|\mu,\theta \rangle&=&V_B^{\dagger}(\theta)|\mu
\rangle,\,\,|\eta,\theta\rangle=V_B^{\dagger}(\theta)|\eta \rangle.
\end{eqnarray}

\section{Entanglement discontinuity}\label{sec:4}

To gain better understanding, and to show how discontinuity comes about, let us first consider a family of TMSS which at the limit of infinite squeezing recover the maximally entangled states discussed above. 

The Wigner representation of a TMSS is given by \cite{bra}
\begin{equation}\label{tmss}
W(\eta,\nu,\mu,\xi)=\frac{4}{\pi^2}{\rm exp}[-2e^{-2r}(\eta^2+\nu^2)-2e^{+2r}(\mu^2+\xi^2)].
\end{equation}
This  Wigner  function  approaches ${C \delta(x_1 - x_2)\,\delta(p_1 + p_2)}$ in the limit of infinite squeezing ${r\rightarrow\infty}$,  corresponding to the original (perfectly correlated i.e. maximally entangled, but unphysical) EPR state (\ref{epr}). While at ${r=0}$ it correspond to two-mode separable coherent states. For all ${r>0}$ this state is entangled. To quantify its entanglement we use here  the measure of `impurity', ${\cal E}$, of the reduced state, that is 
\begin{equation}
{\cal E} =1-\pi^2\int dx_1\,dp_1\left(\int dx_2\,dp_2\, W(x_1,p_1,x_2,p_2)\right)^2.
\end{equation}
This measure is known in the literature as linear or linearized entropy and has
been recently used also as a successful measure of entanglement \cite{purity1,purity2,purity3,purity4,purity5}.

Now consider the rotation $V_B$, Eq.~(\ref{vb}), which  amounts to rotation in phase space \cite{ekert}:
\begin{eqnarray}
\eta\rightarrow\cos\theta\,\eta+\sin\theta\,\mu\,,\\\nonumber
\mu\rightarrow\cos\theta\,\mu-\sin\theta\,\eta\,.
\end{eqnarray}
Thus, by applying this rotation to the TMSS, Eq.~(\ref{tmss}), we parametrize a family of TMSS by an angular variable $\theta$.
The Wigner function of the {$\theta$-parametrized} TMSS is given by
\begin{align}\label{thetatmss}
&W(\eta,\nu,\mu,\xi;\theta)\equiv
W(\cos\theta\,\eta+\sin\theta\,\mu,\nu,\cos\theta\,\mu-\sin\theta\,\eta,\xi)\nonumber\\&=\frac{4}{\pi^2}{\rm exp}\{-2e^{-2r}[(\cos\theta\,\eta+\sin\theta\,\mu)^2
+\nu^2]\nonumber\\&\qquad\qquad\,-2e^{+2r}[(\cos\theta\,\mu-\sin\theta\,\eta)^2+\xi^2]\}.
\end{align}
We first note that at ${\theta=0}$ we obtain the `usual' TMSS given in Eq.~(\ref{tmss}), while at ${\theta=\pi/2}$, \mbox{$W(\eta,\nu,\mu,\xi;\theta=\pi/2)$} represents a {\it separable} two modes each is a squeezed state:
\begin{align}
&W(\eta,\nu,\mu,\xi;\theta=\pi/2)\nonumber\\&=\frac{4}{\pi^2}{\rm exp}[-2 e^{-2 r} (\mu^2+\nu^2)-e^{2 r} (\eta^2+\xi^2)]\\&=\frac{4}{\pi^2}\bigg({\rm exp}[-e^{2 r} x_1^2-e^{-2 r} p_1^2]\bigg)\bigg({\rm exp}[-e^{2 r} x_2^2-e^{-2 r} p_2^2]\bigg).\nonumber
\end{align}
The {$\theta$-parametrized} TMSS,  Eq.~(\ref{thetatmss}), is entangled for all ${r\neq 0}$ and ${\theta\neq\pi/2}$ with,
\begin{eqnarray}
{\cal E}=1-\frac{2}{\sqrt{\left(3+\cosh4r+2\cos2\theta \sinh^22r\right)}}.
\end{eqnarray}

\begin{figure}[ht]
\epsfxsize=.40\textwidth
\centerline{\epsffile{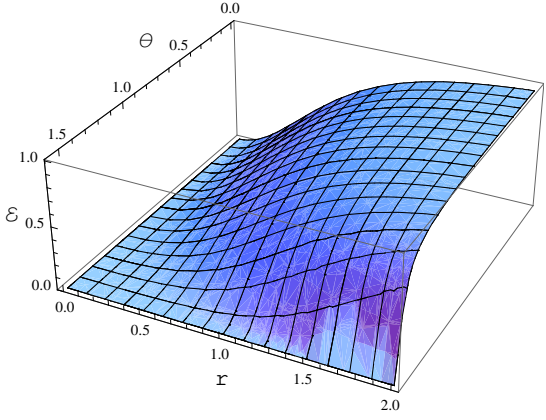}}
\caption{The entanglement, ${\cal E}$, of the {$\theta$-parametrized} TMSS as a function of $r$ and $\theta$. As $r$ increases, the change of the entanglement become more abrupt at the vicinity of ${\theta=\pi/2}$.}
\label{fig1}
\end{figure}

In Fig.~(\ref{fig1}), the entanglement ${\cal E}$ of the {$\theta$-parametrized} TMSS is plotted as a function of ${0\leq r\leq 2}$ and  ${0\leq\theta\leq\pi/2}$. As $r$ increases, the change of the entanglement becomes more abrupt in the vicinity of ${\theta=\pi/2}$. 

\begin{figure}[ht]
\epsfxsize=.50\textwidth
\centerline{\epsffile{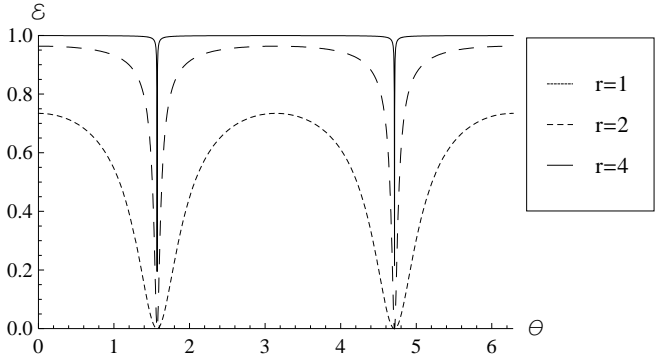}}
\caption{${\cal E}$ as a function of ${0\leq\theta\leq 2\pi}$, for different $r$ values. For ${r\rightarrow\infty}$ the change of ${\cal E}$ is discontinuous.}
\label{fig2}
\end{figure}
Indeed, in Fig.~(\ref{fig2}), we plot ${\cal E}$ as a function of   ${0\leq\theta\leq 2\pi}$ for different $r$ values. We find that for large squeezing the state is (almost) maximally entangled(${\cal E}=1)$ for all ${\theta\neq\pi/2}$ and ${\theta\neq 3\pi/2}$, and is (exactly) product state for ${\theta=\pi/2}$ and ${\theta= 3\pi/2}$.

\begin{figure}[ht]
\epsfxsize=.50\textwidth
\centerline{\epsffile{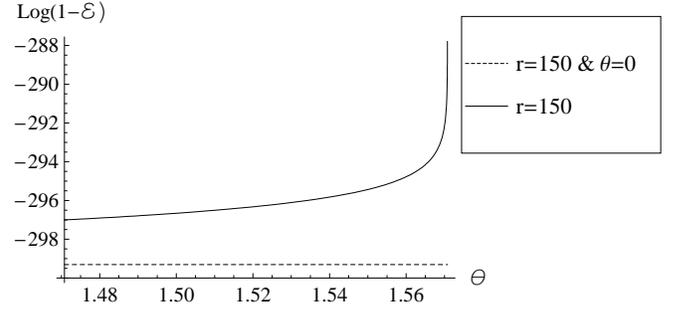}}
\caption{Entanglement, ${\log(1-{\cal E})}$, as a function of ${\pi/2-\epsilon\leq\theta\leq 2\pi}$, with ${\epsilon=0.1}$ for different $r$ values. For ${r\rightarrow\infty}$ the change of ${\cal E}$ is discontinuous.}
\label{fig3}
\end{figure}

To have a closer look at how the entanglement scales in the vicinity of ${\theta= \pi/2}$, we plot, in Fig.~(\ref{fig3}), ${\log(1-{\cal E})}$ as a function of   $\theta$ for different $r$ values. We find that for large enough squeezing the entanglement rapidly drops from maximal value to minimal. This behavior indicates the presence a true discontinuity in the limit ${r\rightarrow\infty}$.

Finally, we show that at the limit ${r\rightarrow\infty}$, the entanglement of the {$\theta$-parametrized} state is maximal (${\cal E}=1$) for all ${\theta\neq \pi/2}$, and the state is separable (${\cal E}=0$) for ${\theta = \pi/2}$. Although in that limit the state is unphysical, the mathematical discontinuity which appears at the limit of a {\it physical} state, enables us to identify this unexpected property of entanglement. At the limit ${r\rightarrow\infty}$ the {$\theta$-parametrized} state ${|\xi\rangle|\mu,\theta \rangle}$, is an eigenstate of $\hat\xi$ and ${(-\sin\theta \,\hat{\eta}+\cos\theta \,\hat{\mu})}$, defined in Eqs.~(\ref{ximu}) and (\ref{EntBases}).
First we note that for ${\theta=\frac{\pi}{2}}$ this is a product state since (c.f., Eq.~(\ref{EntBases})),
\begin{align}\label{product}
&|\xi\rangle|\mu,\frac{\pi}{2}\rangle=|\xi\rangle|-\eta\rangle\nonumber \\&=\int dx_1 dx_2\; \delta\left(\frac{x_1-x_2}{\sqrt
2}-\xi\right)\delta\left(\frac{x_1+x_2}{\sqrt 2}+\eta \right)|x_1,x_2\rangle\nonumber \\&=\int dx_1\;\delta\left(x_1-\frac{\xi-\eta}{\sqrt2}\right)|x_1\rangle\int dx_2\; \delta\left(x_2+\frac{\xi+\eta}{\sqrt2}\right)|x_2\rangle,
\end{align}

Now we argue that for ${0\le\theta <\frac{\pi}{2}}$
the state is {\it maximally} entangled in as much as (i) partial tracing with respect
to one coordinate gives the state of the other coordinate to be proportional to unity.
And (ii), the state when considered within a Schmidt-like expansion involves diagonal
pairing all with equal probability.

We begin by taking the partial trace of an off-diagonal form 
${|\xi\rangle|\mu,\theta\rangle \langle \xi'|\langle
\mu',\theta|}$,
\begin{align}
&\int dx'_1 \langle x'_1|\xi\rangle|\mu,\theta\rangle \langle \xi'|\langle
\mu',\theta|x'_1\rangle \\\nonumber&= \int dx'_1\langle
x'_1| \int d\eta
d{\eta}'d\bar{\eta}d\bar{\eta}'\\\nonumber&\times|\xi,\eta\rangle\langle\eta|\bar{\eta},\theta\rangle\langle\bar{\eta},\theta|\mu,\theta
\rangle\langle\mu',\theta|\bar{\eta}',\theta\rangle
\langle\bar{\eta}',\theta|\eta'\rangle\langle\eta',\xi'|x'_1\rangle .
\end{align}
The various matrix elements are given by,
\begin{align}
\langle x_1|\xi,\eta\rangle\; &=\delta\left(x_1-\frac{\eta+\xi}{\sqrt2}\right)
\Big|x_2=\frac{\eta-\xi}{\sqrt2}\Big\rangle,\nonumber\\
\langle\eta|\bar{\eta},\theta\rangle\;\;\; &=\frac1{\sqrt{2\pi\sin\theta}}e^{-\frac{i}{2\sin\theta}\big[(\eta^2+{\bar{\eta}}^2)\cos\theta-2\eta\bar{\eta}\big]},
\nonumber\\
\langle\bar{\eta},\theta|\mu,\theta\rangle&=\frac{1}{\sqrt{2\pi}}e^{i\bar{\eta}\mu}.
\end{align}
Evaluating the integral we get, 
\begin{align}\label{int2}
&\int d\eta\, e^{\frac{2i\eta }{\sin2\theta}(\Delta_{\xi} \sin^{2}\theta+\Delta_{\mu}
\sin\theta)}\nonumber\\
  &\times e^{\frac{i}{\sin2\theta}\Big[\{\Delta_{\xi}-(\mu+\mu')\sin\theta \}(\Delta_{\xi}+\Delta_{\mu}\sin\theta)-\Delta^{2}_{\xi}\cos^2\theta\Big]}\nonumber\\
  &\times\Big|x_2=\frac{\eta-\xi}{\sqrt2}\Big\rangle\Big\langle x_2=\frac{\eta-\xi+2\Delta_{\xi}}{\sqrt2}\Big|.
\end{align}
Setting ${\Delta_{\xi}=\Delta_{\mu}=0}$ we obtain,
\begin{align}\label{unit}
&\int dx'_1 \langle x'_1|\xi\rangle|\mu,\theta\rangle \langle \xi'|\langle
\mu',\theta|x'_1\rangle \\\nonumber&=
\frac{\sqrt2}{2\pi \cos\theta}\int{d\eta
\Big|x_2=\frac{(\eta-\xi)}{\sqrt2}\Big\rangle\Big\langle x_2=\frac{(\eta-\xi)}{\sqrt2}\Big|}\\\nonumber&=
\frac{1}{\pi \cos\theta}\mathbb{I}_2.
\end{align}
One may readily check the result, Eq.~(\ref{int2}), by integrating over the second variable, $x_2$ to get ${\delta(\xi-\xi')\delta(\mu-\mu')}$ assuring its proper normalization.
The result, Eq. (\ref{unit}), implies that for $0\le\theta<\frac{\pi}{2}$ the state,
$|\xi\rangle|\mu,\theta\rangle$ is maximally entangled. This can also be seen
directly by calculating the $x$ representation of the state and noting that it is of
the same form as the EPR state, i.e. its Schmidt decomposition contains all the
states paired with coefficients of equal magnitude \cite{peres}:
\begin{align}
&|\xi\rangle|\mu,\theta\rangle\\\nonumber&=\int dx_1dx_2|x_1,x_2\rangle \langle
x_1,x_2|\int d\eta d\bar{\eta}\,|\xi,\eta\rangle\langle\eta|\bar{\eta},\theta\rangle
\langle\bar{\eta},\theta|\mu,\theta\rangle \\\nonumber
    &=\frac{\sqrt2}{2\pi\cos\theta}e^{\frac{i\mu}{2\cos
    \theta}\big(2\xi-\mu \sin\theta\big)}
    \int dxe^{\frac{\sqrt2 i x\mu}{\cos\theta}}
    |x\rangle|x-\sqrt2 \xi\rangle.
\end{align}
This is a maximally entangled state for ${0\le\theta<\frac{\pi}{2}}$ while a product state for ${\theta=\frac{\pi}{2}}$ (c.f., Eq	~.(\ref{product})).  We interpret this to mean that entanglement is discontinuous.\vspace{0.3cm}

\section {Conclusions and Remarks}\label{sec:conc}

Single particle mutual unbiased bases labelled by an angle $\theta$ was considered. This labelling was then used to define a  set of two-particle states which we called {$\theta$-parametrized} TMSS. Within this set, we identified a class of states which are almost maximally entangled for ${0\le \theta < \frac{\pi}{2}}$ and  product states for ${\theta=\frac{\pi}{2}}$. This unique class of states may be used to study the power of entangling operation in phase space and has a direct manifestation in (nonlinear) quantum optics. One could use, for example, Bell's inequality for continuous variable systems to demonstrate this feature. It was shown in Ref.~\cite{optbell} how to construct an optimal Bell's inequality for continuous variable Gaussian states such as the {$\theta$-parametrized} TMSS. Then, using the experimental scheme proposed in \cite{experiment} one could test the violation of the suitable constructed Bell's inequality by the {$\theta$-parametrized} TMSS. In this proposed scheme the {$\theta$-parametrized} TMSS is realized by a correlated state of light, where the correlation refers to two spatially separated modes of the electromagnetic field, and $\theta$ is related to a phase of a classical reference field. Next, a photon counting experiment will lead directly to a measurement that is described by the Wigner function. Indeed, it was shown \cite{experiment} that these functions are given by joint photon count correlations and as such can be used to test local realism in the form of Bell's inequalities.  We expect that at the for ${\theta\neq\pi/2}$ and for large squeezing $r$ the violation will be almost maximal, i.e. close to $2\sqrt2$, while at the vicinity of  ${\theta=\pi/2}$ there will be a rapid decrease in the violation. This change become more abrupt as $r$ increases. 
\vfill
\begin{acknowledgments}
Informative discussions with our colleagues Professors J. Zak, A.
Mann and O. Kenneth are gratefully acknowledged. FCK acknowledges financial support from NSERCC. MR thanks the Theoretical Physics Institute for their hospitality. Centre for Quantum Technologies is a Research Centre of Excellence funded by Ministry of Education and National Research Foundation of Singapore.
\end{acknowledgments}

\end{document}